\documentclass[12pt]{article}
\usepackage{epsfig,amssymb}
\begin{document}

%
\catcode`\@=11
\@addtoreset{equation}{section}
\def\theequation{\thesection.\arabic{equation}}
\catcode`\@=12
\newcommand{\be}{ \begin{equation}}
\newcommand{\ee}{ \end{equation}  }
\newcommand{\bea}{ \begin{eqnarray}}
\newcommand{\eea}{ \end{eqnarray}  }
\newcommand{\bi}{\bibitem}
\newcommand{\rd}{ \mbox{\rm d} }
\newcommand{\rD}{ \mbox{\rm D} }
\newcommand{\re}{ \mbox{\rm e} }
\newcommand{\rO}{ \mbox{\rm O} }
\newcommand{\erf}{\mbox{\rm erf}}
\newcommand{\diag}{\mbox{\rm diag}}

\renewcommand{\floatpagefraction}{0.8}

\def\I{\cite{schro}}
\def\del{\partial}
\def\SF{Schr\"odinger functional }
\def\cb{\bar{c}}
\def\q{\tilde{q}}
\def\c{\tilde{c}}

\def\rmd{{\rm d}}
\def\rmD{{\rm D}}
\def\rme{{\rm e}}
\def\rmO{{\rm O}}
\def\tr{{\rm tr}}

 
\def\gms{g_{\ms}}
\def\gmsbar{g_{\msbar}}
\def\gbar{\bar{g}}
\def\gbarms{\gbar_{\ms}}
\def\gbarmom{\gbar_{\rm mom}}
\def\ms{{\rm MS}}
\def\msbar{{\rm \overline{MS\kern-0.14em}\kern0.14em}}
\def\lat{{\rm lat}}
\def\glat{g_{\lat}}
\def\gSF{g_{\rm SF}}

\def\alphabar{\alpha}
\def\alphasf{\alpha_{\rm SF}}
\def\alphat{\tilde{\alpha}_0}

 
\def\SU{{\rm SU}(N)}
\def\SUtwo{{\rm SU}(2)}
\def\SUthree{{\rm SU}(3)}
\def\su{{\rm su}(N)}
\def\sutwo{{\rm su}(2)}
\def\suthree{{\rm su}(3)}
\def\pauli#1{\tau^{#1}}
\def\Ad{{\rm Ad}\,}

\def\rf#1{{#1}}

\newcommand{\nn}{\nonumber}
\newcommand{\phd}{\phi^{\dagger}}
\begin{titlepage}
\vspace*{-2.0cm}

\begin{flushright} DESY 00-104 \end{flushright}\vspace*{-.5cm}
\begin{flushright} HU-EP-00/30 \end{flushright}\vspace*{-.5cm}
\begin{flushright} NIC/DESY-00-002\end{flushright}\vspace*{-.5cm}
\begin{flushright} Bicocca-FT-00-12 \end{flushright}
\vspace*{0.5cm}
\begin{center}
{\LARGE Comparative Benchmarks\\ of full QCD Algorithms}
\vskip 1 cm
\vbox{
\centerline{
\epsfxsize=2.5 true cm
\epsfbox{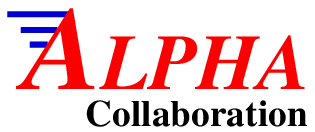}}
}
\vskip 0.5 cm

%
{\large Roberto Frezzotti\\
Dipt. di Fisica, Univ. di Milano Bicocca\\
Via Celoria 16, I-20133 Milano, Italy\\}
\vspace{.5cm}
{\large Martin Hasenbusch, Ulli Wolff\\
Institut f\"ur Physik, Humboldt Universit\"at\\
Invalidenstr. 110, D-10099 Berlin, Germany\\}
\vspace{.5cm}
{\large Jochen Heitger\\
DESY\\
Platanenallee 6, D-15738 Zeuthen, Germany\\}
\vspace{.5cm}
{\large Karl Jansen\\
CERN, Theory Division\\
CH-1211 Geneve 23, Switzerland}
\end{center}
\vspace{.5cm}
\thispagestyle{empty}
\begin{abstract}\normalsize
We report performance benchmarks for several algorithms
that we have used to simulate the Schr\"odinger functional
with two flavors of dynamical quarks. They include hybrid and
polynomial hybrid Monte Carlo with preconditioning.
An appendix describes a method to deal with autocorrelations
for nonlinear functions of primary observables as
they are met here due to reweighting.

\end{abstract}

\end{titlepage}

\section{Introduction}

In the past years the ALPHA Collaboration has pursued the goal
to reliably compute the QCD gauge coupling at high energy in terms
of non-perturbative low energy parameters. The concomitant necessity 
to deal with a large energy ratio in the continuum limit was solved
by a breakup into recursive steps. 
Here one employs finite size
rescaling by repeated factors of two and extrapolates to the
continuum each step by itself. 
By a combination of theoretical reasoning and numerical tests the
\SF was determined as a particularly convenient framework
for this purpose.
The programme has been completed for the quenched
approximation, see refs.~\cite{REV1,REV2} for reviews of the approach
and ref.~\cite{QUENCHEDDATA} for a summary of
data. First tests with a non vanishing flavour number 
have been reported \cite{JURI}.  As is well-known, by the inclusion
of dynamical quarks the numerical cost is boosted by a large factor.
The importance of algorithmic optimization can hence hardly be
overestimated. The finite size technique with the \SF
--- beside its uses for QCD physics ---
offers the possibility of an investigation
of the lattice spacing dependence of the performance of fermion
algorithms with all physical scales  held fixed. Here we report
on such results for several algorithms. 

The \SF can be regarded as the free energy $\Gamma$ of QCD in a finite volume
$L^3 \times T$,
\be
\exp(-\Gamma)=\int D[U]D[\overline{\psi}]D[\psi]
\exp(-S[U,\overline{\psi},\psi]).
\label{SFdef}
\ee
The action $S$ consists of the usual plaquette action for SU(3) gauge fields
$U$ and two degenerate flavours of clover-improved Wilson fermions.
The box is periodic in space, and fixed
gluon potentials
and vanishing quark fields\footnote{
Non-vanishing quark sources are also possible but will not be needed here.
} 
are prescribed
on the temporal boundaries.
The boundary potentials $C$ and $C'$ at $x_0=0$ and $x_0=T$ are specified
in terms of the scale $L$ and dimensionless parameters, one of which
is called $\eta$ and is kept variable.
A convenient
practical choice with $T=L$, 
introduced as point ``A'' in \cite{SFA}, is used throughout.
Quark fields are periodic in space up to a phase $\theta=\pi/5$.
An Abelian background field is induced which can be varied by changing $\eta$.
The response to such an infinitesimal variation is used to define
the renormalized coupling
\be
\gSF^2(L)=\left.\frac{\del\Gamma_0/\del\eta}{\del\Gamma/\del\eta}\right|_{\eta=0}\; ,
\ee
where $\Gamma_0/g_0^2$ is the tree level value of $\Gamma$ for bare
coupling $g_0$. For a lattice realization we now have to choose values
for $L/a$, $g_0$ and bare quark mass $m_0$ as well as coefficients for
the improvement terms in the action. We take the latter as 
smooth functions
of $g_0$ either by a perturbative expression or by a non-perturbative fit
\cite{NPIMPR}. The mass $m_0$ is fixed by demanding zero PCAC-mass \cite{PCAC}.
Hence we may approach the continuum limit by a sequence
of lattices with growing $L/a$ and $g_0$ adjusted to maintain a fixed 
value $\gSF$.
Conceptually this is exactly the same situation
as in our quenched computations.
The regularizing lattice spacing $a$ varies while renormalized physics is
held fixed. It is on such `trajectories', that we study algorithm performance.

Our most extensive simulations of the $\rO(a)$ improved \SF
have been conducted with the well-known
hybrid Monte Carlo method (HMC) \cite{HMC1}. In our implementation we
took advantage of preconditioning and the refinement proposed
in \cite{SEXWEIN}. It amounts to the introduction of two different step sizes
for fermion-gluon and gluonic self-couplings
in an approximately optimal proportion depending on their relative computational
cost. In other long runs we applied the polynomial
hybrid Monte Carlo (PHMC) \cite{deFor,PHMC1}. Here, as for the multiboson technique
\cite{MBML}, an approximately inverting polynomial of the Dirac operator
is used to bosonize the theory. In the multiboson proposal the resulting action is
represented by many boson fields with nearest neighbour couplings.
\rf{For unimproved Wilson fermions
finite step-size updates are employed, which however become impractical
when the clover term is included -- the case on that we concentrate here. 
A further disadvantage is the additional slowing down due to collective
effects of the many bosons \cite{Beat}.}
With PHMC the operator
polynomial is employed to construct a non-local Gaussian action for 
only one boson field which
is simulated by HMC. The imperfection of the polynomial can
be corrected by an acceptance or reweighting step. 
Some results are reported which have been obtained by a recently proposed multi-level
Metropolis procedure (MLM) \cite{MLMH}. 
Further details of the various algorithms will
be given below.

\section{Algorithms in this study}
In this section we briefly describe our implementations
of fermion Monte Carlo algorithms as they are benchmarked in this study.
With each of them the goal is the inclusion of effects of the
weight factor $\det(Q)^2$ which arises from integrating two degenerate
flavours 
of quarks
out of (\ref{SFdef}),
\be
\exp(-\Gamma)=\int D[U]
\exp(-S_{\rm gauge}[U])\, \det(Q)^2.
\label{SFdefQ}
\ee
Here the hermitian operator $Q$ for Sheikholeslami-Wohlert improved
Wilson quarks has the structure
\be
Q = c_0 \gamma_5 M; \quad M=1-T-H.
\label{QMdef}
\ee
The constant $c_0$ is chosen to contain the eigenvalues of $Q$
in the interior of the interval $(-1,1)$. 
The matrix $M$ contains 
nearest neighbour hopping terms in $H$ and the clover term in $T$, which
is diagonal with respect to the lattice index.
The detailed form of these components, including
boundary improvement,  can for instance be found in 
ref.~\cite{Qdefinition}.

\subsection{Hybrid Monte Carlo}

The HMC method \cite{HMC1} has so far been the most popular
fermion algorithm for QCD. 
In choosing a trajectory length of unity
we followed the general experience that this is close to optimal.
In \cite{Bernd} this was confirmed for the quenched Schr\"odinger
functional, and a test with dynamical fermions at $L/a=8$
showed an almost doubling of computational costs 
for $g_{\rm SF}^2$ as we lowered
the trajectory length to one half. 
We
reduced discretization errors by the multiple time scale method
proposed in ref. \cite{SEXWEIN} taking the version given there in 
eq.~(6.7) with $n=4$. A test of the performance 
gain of the above integration scheme 
in practical simulations was performed in 
\cite{JaLi95}, where it was demonstrated that a substantial gain
is achieved as
compared to a standard leap-frog integrator. The value of $n$ was
not varied any further in this study.

As an essential sophistication we made use of two different forms of
preconditioning. Both rely on our ability to factor out of $Q$
matrix factors which on the one hand are easy to invert and on the 
other hand capture a part of its spectral variation to leave us with a better
conditioned remaining factor.  For even-odd preconditioning
we exhibit the block structure of $M$ with respect to even (e) and
odd (o) lattice sites and factorize
\be
M = 
\left( 
\begin{array}{cc} 
M_{\rm ee} & M_{\rm eo} \\
M_{\rm oe} & M_{\rm oo} 
\end{array}
\right) = 
\left( 
\begin{array}{cc} 
M_{\rm ee} & 0 \\
M_{\rm oe} & 1 
\end{array}
\right) 
\times 
\left( 
\begin{array}{cc} 
1 & M_{\rm ee}^{-1} M_{\rm eo} \\
0 & M_{\rm oo} - M_{\rm oe}  M_{\rm ee}^{-1} M_{\rm eo}
\end{array}
\right),
\label{factoreo}
\ee
where the left (lower) block-triangular factor and the block-diagonal
$M_{\rm ee}$ are easy to invert. This factorization can now be used
in a two-fold way. If the original $Q$ under the determinant in (\ref{SFdefQ})
is plugged into the HMC algorithm
we have to continuously solve linear systems with coefficient matrices
given by $Q$. With (\ref{factoreo}) these can
be transformed into better conditioned systems 
with accelerated iterative inversion of
\be
\hat{Q} = \tilde{c}_0 \gamma_5 
(M_{\rm oo} - M_{\rm oe}  M_{\rm ee}^{-1} M_{\rm eo}).
\label{Qhat}
\ee
The constant $\tilde{c}_0$ is again used to normalize the spectrum 
of $\hat{Q}$.
On the other hand we may also
conclude from (\ref{factoreo}) that up to irrelevant constant factors
the relation
\be
\det(Q) \propto \det(M_{\rm ee}) \det(\hat{Q})
\ee
holds.
Now $\hat{Q}$ enters into the HMC and leads to a different 
Monte Carlo dynamics, which also takes $\det(M_{\rm ee})$  into account.
When we refer to even-odd preconditioning in this paper, 
this second variant will
always be meant. Further details on our implementation of HMC may be found
in \cite{JaLi96}. \rf{As we only have to invert the squared
operator $\hat{Q}^2$ we use the conjugate
gradient  method (CG), which was found to be close to optimal in this case.}

For SSOR preconditioning  a different factorization of $M$ based on factors
triangular with respect to a lexicographic ordering of lattice sites is used
\cite{ssor,ssor1,ssor2}. 
Due to its complexity, \rf{in particular if the clover term is included,}
this has  to our knowledge only been used
to accelerate linear systems and was for that purpose reported to be 
superior
over even-odd preconditioning if combined
with the BiCGstab \cite{bicg} inversion algorithm \rf{for the preconditioned
$M$ and $M^\dagger$.}
In the following SSOR will refer to such an implementation.
\rf{For the unimproved case, a simplified form of SSOR preconditioning
(ILU) was implemented under the determinant with very positive 
results~\cite{deFor}.}

We compared the performance of our two HMC program versions on our largest
lattice, i.e. $12^4$ at $\beta=9.5$. 
In solving the linear systems with the respective preconditioned operators
we confirmed that in terms of \rf{operations associated with applying
these operators to fields,} the BiCGstab
algorithm with SSOR preconditioning outperforms the CG algorithm 
with even-odd preconditioning by a factor of about 1.6.
Part of this advantage is however lost in terms of CPU time,
because on our Alenia Quadrics (APE) machines inner products 
are relatively expensive. Since in the BiCGstab algorithm inner products
and linear combinations are much more frequent than in the CG algorithm, 
this is a non-negligible overhead.
The overall advantage that we find for the even-odd version
derives however from the different operators under the
determinant.
A clear sign
of this is the behaviour of the acceptance rate in both cases. 
While for the \rf{even-odd
preconditioned determinant} 
we could obtain an acceptance rate of $91\%$ with a
step size of $\Delta\tau=0.08$, for $\det(Q^2)$
it went down to $75\%$ already at a step size of $\Delta\tau=0.07$.

In principle, one could also conceive of the following combination
yet untested by us. One uses the even-odd preconditioned determinant
and, when linear systems with $\hat{Q}$ have to be solved, one
transforms them to the SSOR preconditioned form, solves, and
translates back. It is unclear at present, whether the overhead
still leaves this variant profitable.

\subsection{Polynomial Hybrid Monte Carlo}

We recall here some basics of the PHMC algorithm.
For technical details the reader
is referred to refs. \cite{PHMC2,PHMC3}.
In the PHMC algorithm the inverse of $\hat{Q}^2$ is approximately 
computed 
by 
a suitable \cite{MBML} Chebyshev polynomial  of degree $n$,
\be
\hat{Q}^{-2} \approx P_{n,\epsilon}(\hat{Q}^2).
\ee
Defining the relative deviation
\be
R_{n,\epsilon}(\lambda) = \lambda P_{n,\epsilon}(\lambda) -1,
\ee
the inversion error
for eigenvalues $\lambda\in [\epsilon,1]$ of $\hat{Q}^2$
is bounded by
\begin{equation} \label{accuracy}
\delta = \sup_{\lambda} |R_{n,\epsilon}(\lambda)| = 2
\left(\frac{1-\sqrt{\epsilon}}{1+\sqrt{\epsilon}}\right)^{n+1}
\; .
\end{equation}
For a given degree $n$ the free parameter $\epsilon$ in $P_{n,\epsilon}$
allows to trade between approximation range and accuracy.
For eigenvalues $\lambda < \epsilon$ the error
monotonically moves from $R_{n,\epsilon}(\epsilon) = -\delta$
to 
$R_{n,\epsilon}(0) = -1$.
With the help of $P_{n,\epsilon}$ we represent the determinant
by a bosonic spinor field (pseudofermion) $\phi$
\be
\det(\hat{Q}^2) = \int D [\phi]  D [\phd] \, \exp(-S_P) \,  W
\label{detrep}
\ee
with the Gaussian action
\be
S_P = \phd P_{n,\epsilon}(\hat{Q}^2[U]) \phi
\label{PFaction}
\ee
and the remainder
\be
W=\det(\hat{Q}^2 P_{n,\epsilon}(\hat{Q}^2))
\ee
rendering (\ref{detrep}) exact.
As long as the spectrum
of $\hat{Q}^2$ is in the approximation range $[\epsilon,1]$, 
$W$ is a small correction
close to one.
Expectation values 
in the full QCD ensemble are now given
by reweighting with $W$ as
\be
\langle {\cal O} \rangle = \frac
{\langle {\cal O}W \rangle_P}  
{\langle W \rangle_P} \; ,
\label{true_ave}
\ee
where ${\cal O}$ is some observable 
and the average $\langle \dots \rangle_P$ is taken with the action 
$S_{\rm gauge}+S_P$.
Since $W$ is still given by a determinant,
a straightforward evaluation is hard.
As it is a small correction, however, stochastic 
(unbiased) estimates
should be adequate. For each measurement we construct
an estimator $\overline{W}$ given by
\be
\overline{W} = \frac{1}{N_{\rm corr}} \sum_{i=1}^{N_{\rm corr}}
\exp\left\{\eta_i^\dagger (1-[\hat{Q}^2 P_{n,\epsilon}(\hat{Q}^2)]^{-1}) \eta_i \right\}
\label{DefWbar}
\ee
with independent Gaussian random fields $\eta_i$.
Averaging over
$N_{\rm corr}$ such estimates allows us to reduce and control
the extra noise inflicted here.
The true QCD average is then estimated by eq.~(\ref{true_ave}) 
with $W$ replaced by 
$\overline{W}$. 

The update of the gauge field and the pseudofermionic field
$\phi$ follows the standard HMC pattern with global heatbath for
$\phi$ and molecular dynamics for $U$ including the speedup from
\cite{SEXWEIN} discussed before. This is chosen, because,
in contrast to the
multiboson algorithm \cite{MBML}, finite step size updates for $U$
are impractical here due to the complicated non-local effective action.

At this point
the parameters $n, \epsilon$ and, less importantly, $N_{\rm corr}$
are at our disposal for optimization. For small
eigenvalues the growth of the `inverter'
$P_{n,\epsilon}(\lambda)\sim 1/\lambda$ is
cut off at $\lambda\sim\epsilon$. For the HMC dynamics $\epsilon$ hence, 
in some sense, takes
over the role of the smallest eigenvalue. It was found advantageous \cite{PHMC2,PHMC3}
to choose $\epsilon$ a few times larger than the typical smallest eigenvalue of $\hat{Q}^2$.
This allows us to keep the degree of the polynomial lower
for the same approximation accuracy. 
Configurations with small eigenvalues of $\hat{Q}^2$ are produced more frequently
than they would be with the exact determinantal weight. As the algorithm is still
exact, 
this is precisely compensated by $W$ or respectively $\overline{W}$ giving
smaller weight to the observables evaluated on these configurations.
It should be borne in mind that the unquenched lattice path integral is always well-defined.
The potential problem with nearly `exceptional' configurations is  a statistical one
with rarely sampled large contributions, which is alleviated by our sampling and
reweighting technique.
This is the reason for us to prefer the reweighting correction
over an acceptance step.

There is a special round-off problem for PHMC that we briefly summarize now
with more details available in \cite{PHMC2,BUNK}.
To generate $\phi$
with action (\ref{PFaction}) it is necessary to factorize
\be
P_{n,\epsilon}(\hat{Q}^2) = F_n(\hat{Q})^\dagger F_n(\hat{Q})
\ee
with an $n$'th degree polynomial $F_n$. 
For gauge field updating $U$-derivatives of $S_P$
have to be  taken at fixed $\phi$.
To this end we factorize further
\be
F_n(\hat{Q})\phi = [\sqrt{c_n}(\hat{Q}-r_n)] [\sqrt{c_{n-1}}(\hat{Q}-r_{n-1})] \cdots
[\sqrt{c_1}(\hat{Q}-r_1)]\phi
\label{ordering}
\ee
and store the occurring subproducts to facilitate the force computation.
While the complex roots $r_k$ are determined by $P_{n,\epsilon}$,
the real factors $\sqrt{c_k}$ only serve to prevent the partial products from
growing too large or too small. 
It is known \cite{BUNK} that the evaluation of a high order matrix polynomial
in factorized form is in principle rather susceptible to round-off error.
In particular, the ordering of factors in (\ref{ordering}) is of crucial importance
in this context. In \cite{PHMC2,BUNK} orderings were found
which  make this source of errors negligible for the runs 
with $n$ up to 46 reported in this paper,
even on the 32-bit APE 100 machines.
Further details on tuning the polynomial parameters are deferred to appendix B.

\rf{Another variant of PHMC could be devised by applying an inverting
polynomial to the complex spectrum of 
$\gamma_5 \hat{Q}$  instead of
the real positive $\hat{Q}^2$. A corresponding multiboson algorithm
was investigated in ref.~\cite{NHMB}. 
We shall investigate this method in the near future.}

\subsection{Multi-Level Metropolis Algorithm}

As for the previous algorithms it is
our aim to represent $\det (Q^2)\propto \det (M^\dagger M) $ in a way 
suitable for simulation. Here this will be done in part by the explicit use
of a few terms of the hopping parameter expansion in powers of 
$T+H$ (see eq.~(\ref{QMdef}))
and by integrals over a collection of pseudofermion fields to represent the
remainder.

The series for the logarithm of the determinant is given by
\begin{equation}
\log\det (M) = \tr \ln (M) = - \tr (T+H) - \frac{1}{2}  \tr (T+H)^2
 - \frac{1}{3} \tr (T+H)^3 + \ldots \;.
\end{equation}
In our algorithm, we separate off  the series up to some order $k$. 
In the present work we found it convenient to use $k=3$, since
$\tr (T+H)^i = \tr T^i$ can easily be computed for $i=1,2,3$. At higher orders
also mixed terms would contribute. In order to deal with the remaining terms,
we define 
\begin{equation}
 \tilde M = M \; \exp\left(\sum_{j=1}^k \; \frac{1}{j}\; (T+H)^j\right)
 \;\;\;.
\end{equation}
For the inverse of $\tilde M$ we introduce a hierarchical approximation
by polynomials $P_i$ of order $n_i$ in $T+H$,
\begin{equation}
 \tilde M^{-1} =  \prod_{i=1}^{j}P_i^{r_i} + \rO ((T+H)^{n_j+1}).
\end{equation}
The full required inversion accuracy is reached for the maximal value $j=l$.
This is used with $r_1+r_2+\ldots +r_l$ pseudofermion fields to
derive the representation
\be
\det (\tilde M^\dagger \tilde M) \propto
\int \prod_{i=1}^l \prod_{a=1}^{r_i} D[\phi^{\dagger}_{ia}] \; D[\phi_{ia}]
\; \exp \left( -\sum_{i,a} |P_i \, \phi_{ia}|^2 \right) .
\ee
\rf{The powers $r_i>1$ are advantageous as they lead to a smaller
force on the gauge field which allows larger update steps~\cite{MLMH}.}

Let us summarize the action that is simulated,
\begin{equation}
 S  =  S_{\rm gauge}[U]  +  S_{\rm hop}[U]  + S_{\rm PF}[U,\phi]
\end{equation}
with\footnote{
$\tr T$ is small but nonzero, as we also included boundary
improvement terms in it.
}
\begin{equation}
S_{\rm hop}[U] = \tr T  + \frac{1}{2} \tr T^2
+ \frac{1}{3} \tr T^3
\end{equation}
and
\begin{equation}      
S_{\rm PF}[U,\phi] = \sum_{i,a} |P_i \, \phi_{ia}|^2 \; .
\end{equation}

We now consider $S_{\rm gauge}[U]$ as
a zeroth approximation which is taken into account
in generating finite stepsize
primary update {\it proposals}. All further terms will eventually
be implemented by accept-reject filters. In the first level action
\begin{equation}
 S^{(1)} = S_{\rm gauge} + S_{\rm hop}  + \sum_{a=1}^{r_1} |P_1 \, \phi_{1a}|^2
\end{equation}
the hopping term is included as it would otherwise lead to a single link action too
complex for level zero. All further levels $j>1$ are straightforwardly
given by
\begin{equation}
 S^{(j)} = S^{(j-1)} +   \sum_{a=1}^{r_j} |P_j \, \phi_{ja}|^2
\end{equation}
up to $j=l$.

We now describe the steps of 
the multi-level Metropolis update scheme.
A proposal at level 0 is given by multiple local updates with the
Cabibbo-Marinari heatbath or the overrelaxation algorithm.
One possibility  is 
to randomly select $v$ links to be updated. 
In ref. \cite{MLMH} we found it advantageous to actually update 
several times
a sublattice chosen at random from a set that covers the lattice.
The size of the sublattices is chosen such that a reasonable acceptance at
level one of the algorithm is  achieved.
The whole  proposal
obeys detailed balance with respect to the level zero action as we
reverse the order of link updates with probability 1/2.
The remaining levels now proceed recursively as follows.
Generate a proposal $\tilde U$ by performing $t_j$ update steps at level $j-1$.
Accept  $\tilde U$ as new configuration at level $j$ with the probability
\begin{equation}
 A^{(j)} \;=\; \min \left[1,\exp(-\Delta S^{(j)}[\tilde U] + 
\Delta S^{(j)}[U])\right],
\end{equation}
where $\Delta S^{(j)} = S^{(j)} -  S^{(j-1)} $.

In our implementation, the auxiliary fields $\phi_{ia}$ are kept 
fixed during the update
cycle of the gauge field after they have been generated by a global
heatbath by solving
\begin{equation}
 \phi_{i,a} = P_{i}^{-1} \eta,
\end{equation}
where $\eta$ is a Gaussian random field.

In its present form the MLM algorithm will not be the method of
choice for large lattices. The reason is that its cost
will ultimately grow proportional to the square of the number
of lattice sites. 
This is because the size of the updated blocks cannot
grow while maintaining the acceptance rate and thus
their number is proportional to the volume. Each evaluation of 
$\Delta S^{(j)}$ is however also of order volume in complexity.
On the other hand, as we shall see shortly,
MLM can produce precise results at $L/a=5$ where we have tested it here. 
As it contains interesting elements, for instance being a
finite step-size method for improved dynamical fermions, 
we still found the idea
and the practical test
worth reporting and comparing with other methods here, 
for instance as a basis for further modification.

\section{Benchmarks of algorithmic performance}

\subsection{Our measure of efficiency}

We now define two quantities, $M_{\rm cost}$ and $D_{\rm cost}$,
which allow to compare simulation costs 
for a certain physics output between different algorithms
and parameters.
The first measure is machine dependent
and refers to actual CPU time on
the APE100 line of parallel computers currently in use by 
the ALPHA Collaboration. 
The second one is
machine independent with the number of Dirac operator applications to
a spinor field being our currency. As a target quantity, whose statistical
accuracy is used for weighing costs in either units, we take our coupling
$\gSF^2$.
The precise definitions are
\bea
M_{\rm cost} = &&(\mbox{update time in seconds on machine M}) \nonumber\\
&&\times \;
(\mbox{error of } 1/\gSF^2)^2
\times (4a/T)(4a/L)^3
\label{Mcost}
\eea
and
\be
D_{\rm cost} = (\mbox{number of applications } Q\phi) \times
(\mbox{error of } 1/\gSF^2)^2 \; .
\label{Dcost}
\ee
Since the squared error in both formulas goes down inversely proportional
to the run length, both quantities, extracted from given Monte Carlo simulations,
do not depend on their lengths. The reason for focusing on absolute errors
of $1/\gSF^2$ is as follows. We assume for the {\em purpose of error analysis
only} that the running of $\gSF$ with $L$ has the structure of 1-loop perturbation
theory, $1/\gSF^2 \approx -2 b_0 \log L + {\rm const}$. Then
\be
\frac{\delta L}{L} = \frac{1}{2b_0}\delta (1/\gSF^2) 
\approx 8 \, \delta (1/\gSF^2)
\ee
holds, and we approximately aim at a certain relative scale uncertainty,
independently of the size of $\gSF$.
In (\ref{Mcost}) the trivial growth proportional to the number 
of lattice sites 
is cancelled
such that both quantities scale in the same way.
The reference machine M in this publication 
will always be the smallest 8-node
machine of type Q1. Most of our data actually come from bigger machines
with up to 512 nodes. The costs on these machines
are converted by multiplying naively
by the ratio of nodes, e.g. 512/8. This means, we neglect communication
overheads, which is a small effect on our hardware and implementation.
Note that with our definitions
costs can be meaningfully compared also under trivial (replica) 
parallelization, of which we make extensive use.

While most of the CPU time with dynamical fermions is spent on
applications of the Dirac operator there is also quite some
overhead from other operations, in particular on small lattices.
This was neglected in $D_{\rm cost}$ 
except for the contribution to the gauge field
force in the PHMC algorithm which is proportional to the polynomial
degree $n$. As a consequence, the ratio $D_{\rm cost}/M_{\rm cost}$
varies between 50\% and 80\% of the theoretical value referring 
to
$Q\phi$ operations only.
Applications of $Q$ and $\hat{Q}$ are so \rf{close to each other
both in theoretical complexity and actual CPU time,}
that we neglect their difference.
The SSOR preconditioned operator, on the other hand,
is counted as 4/3 $Q$ operations due
to extra multiplications of the diagonal (clover) part.

A typical run that entered our benchmarks is entry 12k in Table 1.
With $M_{\rm cost} \approx 12$ and $D_{\rm cost} \approx 3000$
we ran a total of 13000 
trajectories on an $L/a=T/a=12$ lattice and achieved 6 \% scale accuracy
in about 6 days on 256 nodes of APE100.

\subsection{Numerical results}
\newcommand{\gbarsq}{\bar{g}^2}
\newcommand{\gbsqSF}{g^2_{{\rm SF}}}
\newcommand{\pacc}{P_{\rm acc}}
\newcommand{\nmd}{N_{\rm md}}
\newcommand{\tauint}{\tau_{\rm int}}
\newcommand{\mcost}{M_{\rm cost}}
\newcommand{\dcost}{D_{\rm cost}}
\begin{table}[tb]
\centering
\begin{tabular}{lcllr@{.}lr@{.}l}
\hline \\[-1.0ex]
set &
$L/a$ &
$\beta$ &
algorithm &
\multicolumn{2}{c}{$\Delta\tau$} &
\multicolumn{2}{c}{$\pacc$} \\[1.0ex] 
\hline \\[-1.0ex]
12a & 12 & 9.5    & HMC (e/o)  & 0&080 & 0&91 \\
12b & 12 & 9.5    & HMC (SSOR) & 0&060 & 0&86 \\
12c & 12 & 9.5    & HMC (SSOR) & 0&070 & 0&75 \\
12d & 12 & 9.5    & PHMC (e/o) & 0&091 & 0&83 \\
12e & 12 & 9.5    & PHMC (e/o) & 0&100 & 0&76 \\
12f & 12 & 8.5    & HMC (e/o)  & 0&070 & 0&93 \\
12g & 12 & 8.5    & PHMC (e/o) & 0&091 & 0&80 \\
12h & 12 & 8.5    & PHMC (e/o) & 0&114 & 0&75 \\
12i & 12 & 7.5    & HMC (e/o)  & 0&075 & 0&89 \\
12j & 12 & 7.5    & PHMC (e/o) & 0&098 & 0&77 \\
12k & 12 & 7.5    & PHMC (e/o) & 0&098 & 0&78 \\[1.0ex]
\hline \\[-1.0ex]
10a & 10 & 9.3884 & HMC (SSOR) & 0&080 & 0&78 \\
10b & 10 & 8.39   & HMC (SSOR) & 0&080 & 0&75 \\
10c & 10 & 7.3619 & HMC (SSOR) & 0&070 & 0&79 \\
10d & 10 & 6.877  & HMC (SSOR) & 0&070 & 0&72 \\
10e & 10 & 6.5    & HMC (SSOR) & 0&060 & 0&80 \\
10f & 10 & 6.0    & HMC (SSOR) & 0&050 & 0&83 \\
10g & 10 & 5.5    & HMC (SSOR) & 0&040 & 0&82 \\[1.0ex]
\hline \\[-1.0ex]
 8a & 8  & 9.2364 & HMC (e/o)  & 0&080 & 0&96 \\
 8b & 8  & 8.2373 & HMC (e/o)  & 0&120 & 0&91 \\
 8c & 8  & 7.2103 & HMC (SSOR) & 0&100 & 0&71 \\[1.0ex] 
\hline \\[-1.0ex]
 6a & 6  & 9.5    & HMC (e/o)  & 0&110 & 0&97 \\  
 6b & 6  & 9.0    & HMC (e/o)  & 0&110 & 0&97 \\
 6c & 6  & 8.5    & HMC (e/o)  & 0&100 & 0&97 \\ 
 6d & 6  & 7.5    & HMC (e/o)  & 0&070 & 0&98 \\[1.0ex] 
\hline \\[-1.0ex]
 5a & 5  & 9.3884 & HMC (SSOR) & 0&120 & 0&92 \\  
 5b & 5  & 7.3619 & HMC (SSOR) & 0&120 & 0&90 \\[1.0ex] 
\hline \\[-1.0ex]
 4a & 4  & 9.2364 & HMC (e/o)  & 0&130 & 0&98 \\  
 4b & 4  & 8.24   & HMC (e/o)  & 0&120 & 0&98 \\ 
 4c & 4  & 7.21   & HMC (e/o)  & 0&200 & 0&93 \\[1.0ex] 
\hline
\end{tabular}
\caption[]{
Summary of the simulated parameter sets, which enter our performance studies
of dynamical fermion algorithms. Quark masses are close to their critical
values.
\label{t_resSets}
}
\end{table}

\begin{table}[tb]
\centering
\begin{tabular}{lr@{.}lr@{.}lccc}
\hline \\[-1.0ex]
set &
\multicolumn{2}{c}{$\gbsqSF$} &
\multicolumn{2}{c}{$\tauint$} &
$t_{\rm update}/[\,s\,]$ &
$\mcost$ &
$\dcost/10^2$ \\[1.0ex]
\hline \\[-1.0ex]
12a & 1&1 & 2&00(16) & 1582 & 15(1)    & 36(3)    \\
12b & 1&1 & 1&98(23) & 2392 & 20(2)    & 30(4)    \\
12c & 1&1 & 2&82(27) & 1773 & 22(2)    & 37(4)    \\
12d & 1&1 & 1&41(10) & 1636 & 12.9(9)   & 30(2)    \\
12e & 1&1 & 1&35(7)  & 1335 & 12.0(6)  & 28(2)    \\
12f & 1&3 & 2&19(12) & 1768 & 15.0(8)  & 38(2)    \\
12g & 1&3 & 1&16(7) & 1518 & 18.3(1.1)   & 44(3)    \\
12h & 1&3 & 2&47(15) & 1103  & 11.6(7)  & 27(2)    \\
12i & 1&7 & 2&35(14) & 1895 & 14.7(9)  & 39(2)    \\
12j & 1&7 & 2&05(14) & 1554 & 12.6(9)   & 32(2)    \\
12k & 1&7 & 1&81(13) & 1371 & 11.6(9)  & 28(2)    \\[1.0ex]
\hline \\[-1.0ex]
10a & 1&1 & 1&95(7)  & 707  & 9.7(4)   & 16.2(6)  \\
10b & 1&3 & 2&35(10) & 719  & 10.7(5)  & 17.5(8)  \\
10c & 1&7 & 2&63(13) & 804  & 11.5(6)  & 18.7(9)  \\
10d & 2&1 & 3&42(20) & 806  & 14.4(8)  & 25(1)    \\
10e & 2&4 & 3&46(21) & 991  & 16(1)    & 28(2)    \\
10f & 3&3 & 3&42(21) & 1274 & 19(1)    & 33(2)    \\
10g & 5&4 & 3&94(22) & 1932 & 30(2)    & 49(3)    \\[1.0ex]
\hline \\[-1.0ex]
 8a & 1&1 & 1&14(4)  & 260  & 4.0(1)   & 8.2(3)   \\
 8b & 1&3 & 1&40(6)  & 183  & 3.0(1)   & 6.2(3)   \\
 8c & 1&7 & 2&40(9)  & 239  & 5.8(2)   & 8.4(3)   \\[1.0ex]
\hline \\[-1.0ex]
 6a & 1&0 & 0&70(2)  & 49   & 1.00(3)  & 1.98(6)  \\
 6b & 1&1 & 0&75(2)  & 49   & 1.04(3)  & 2.09(6)  \\
 6c & 1&2 & 0&80(2)  & 55   & 1.15(3)  & 2.34(6)  \\
 6d & 1&5 & 1&08(2)  & 74   & 2.08(4)  & 4.50(8)  \\[1.0ex]
\hline \\[-1.0ex]
 5a & 1&0 & 0&67(1)  & 22   & 0.70(1)  & 1.03(2)  \\ 
 5b & 1&5 & 1&02(3)  & 22   & 0.83(2)  & 1.26(4)  \\[1.0ex]
\hline \\[-1.0ex]
 4a & 1&0 & 0&53(1)  & 6    & 0.268(5) & 0.428(8) \\
 4b & 1&2 & 0&57(1)  & 7    & 0.271(5) & 0.445(8) \\
 4c & 1&5 & 0&74(1)  & 5    & 0.222(3) & 0.397(5) \\[1.0ex]
\hline
\end{tabular}
\caption[]{
Cost estimates for the
simulations in the parameter sets specified in Table~\ref{t_resSets}.
Precise results and analysis on the running of $\gbsqSF$ will appear in
\cite{NF2}.
\label{t_resCost}
}
\end{table}

In Table 1 we list the most important parameters
for a subset of our \rf{HMC and PHMC runs} performed
to investigate the QCD running gauge coupling with two
massless flavours\footnote{
The physical implications of these data 
will be analyzed in a separate paper \cite{NF2} while
here we focus on algorithmic aspects.}.
The fourth column indicates which of the algorithms
discussed before was used. This table has to be read
in conjunction with Table 2, where the measured values
for $\mcost$ and $\dcost$ are given. \rf{Results for $\dcost$ of MLM follow
in Table~4 below.}
Error estimates for the costs
stem from the error of $\tauint$ 
of $g^{-2}_{{\rm SF}}$ which is
determined by the method of Appendix A. Its value here refers to
a unit given by complete update cycles (trajectories).
The update time in seconds for one such cycle
is normalized to the APE100-Q1 as discussed
with the definition of $\mcost$. 
\begin{figure}
\vspace{0.0cm}
\begin{center}
\psfig{file=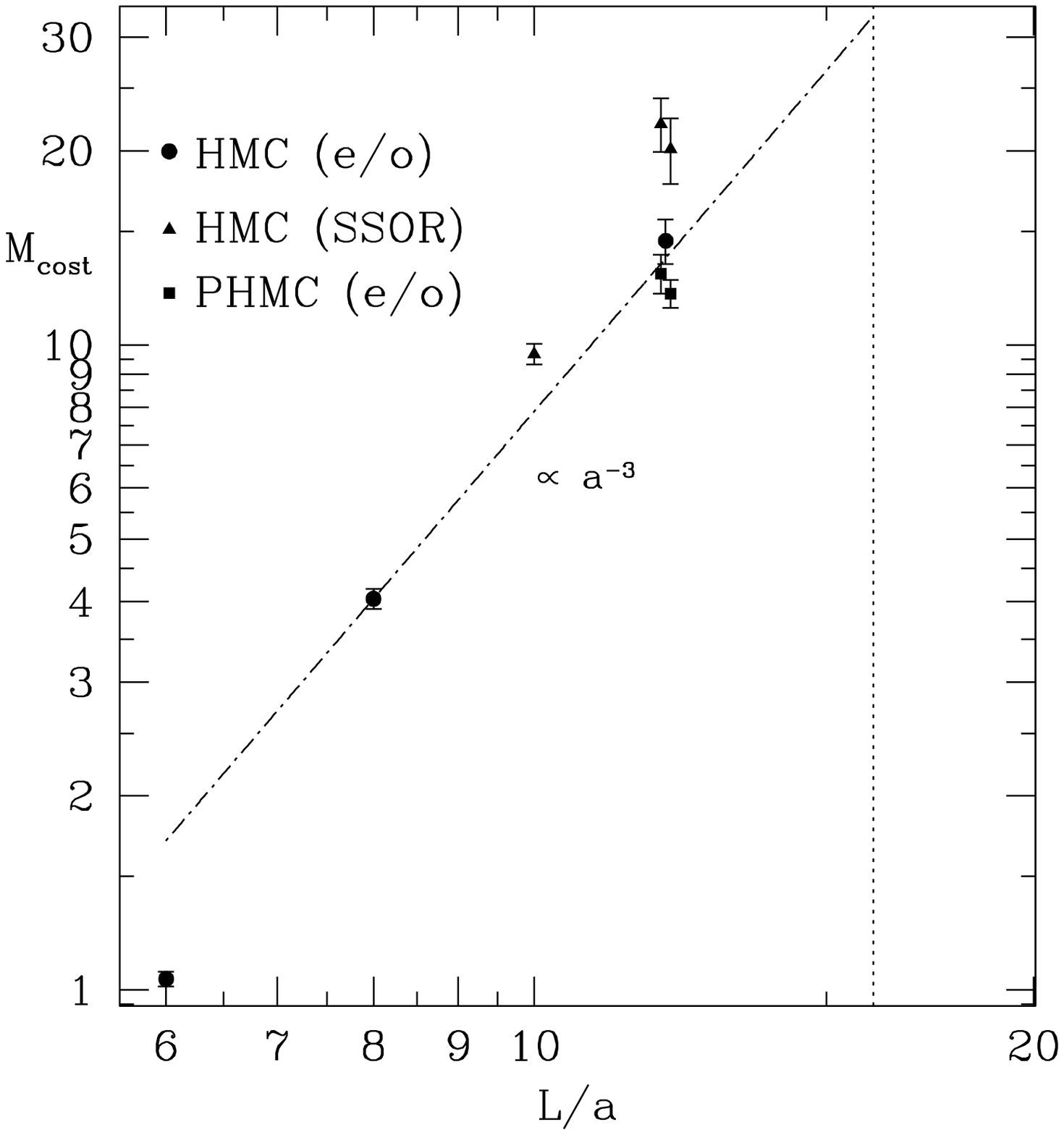, %
width=14cm,height=12cm}
\end{center}
\caption{Measured values of $\mcost$ for runs with
constant physics fixed by $g^2_{\rm SF} \approx 1.1 $ 
and vanishing quark mass.}
\end{figure}
In Fig.~1 we plot $\mcost$ against $L/a$ for all our runs at a fixed
scale $L$ in physical units that is implicitly determined by the
condition $g^2_{\rm SF} \approx 1.1$. 
\rf{The line corresponding to a growth proportional to $a^{-3}$
is shown as a reference and roughly represents the rise of the data.}
This combines effects of a growing
variance of our observable in the Schr\"odinger functional
for $g^2_{\rm SF}$ and of critical slowing
down. The latter accounts for about two powers of $1/a$. 
In Fig.~2 the number of conjugate gradient iterations is shown
for our even-odd preconditioned HMC runs.
\begin{figure}
\vspace{0.0cm}
\begin{center}
\psfig{file=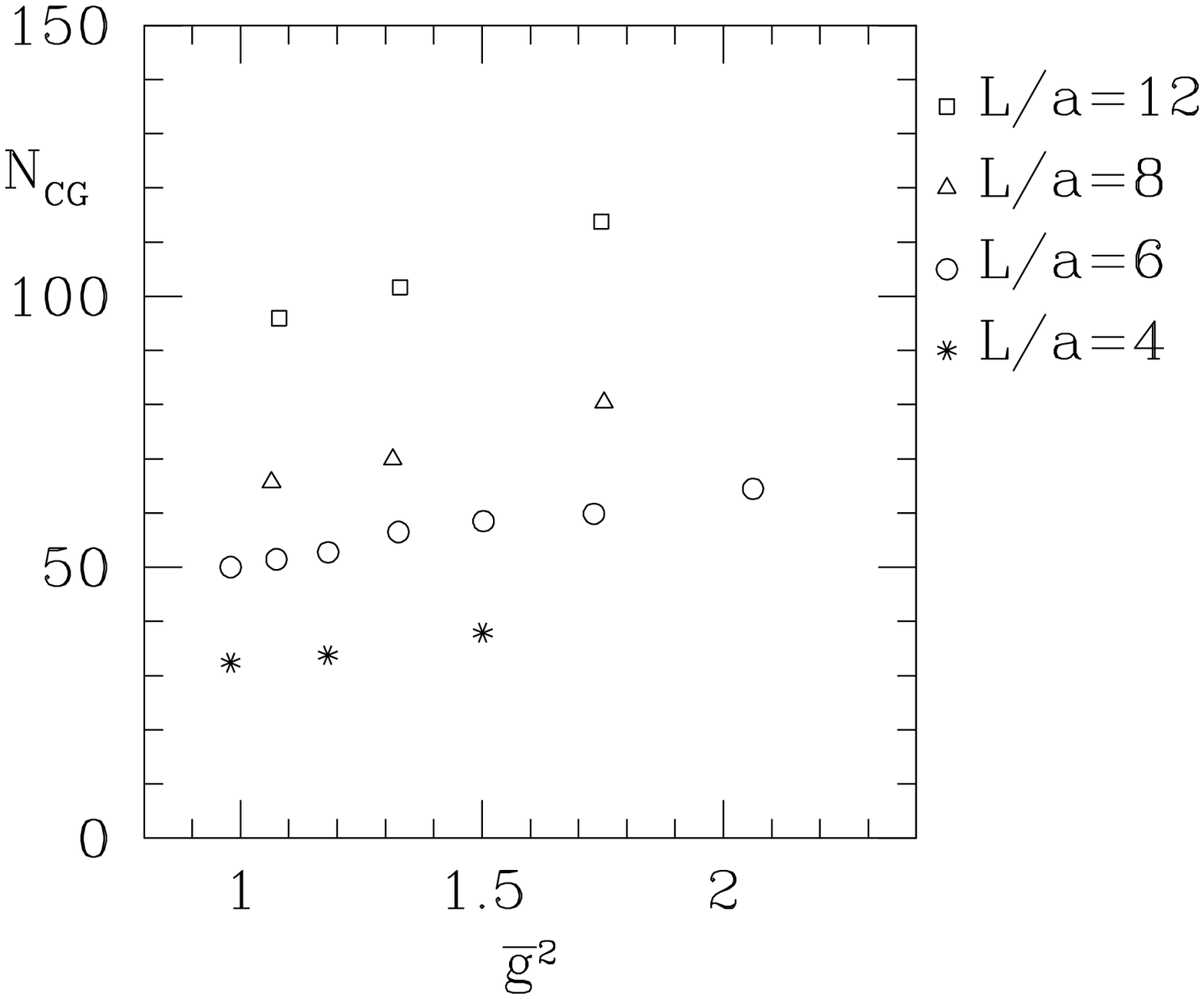, %
width=11cm,height=11cm}
\end{center}
\caption{ \label{fig:ncg} Average number of iterations of 
the conjugate gradient algorithm in even-odd preconditioned HMC runs.}
\end{figure}
At least for smaller $g^2_{\rm SF}$ there is an approximately
linear growth with $L/a$ which contributes one power to critical
slowing down. 
\rf{This is the expected behaviour since $1/L$ is the infrared cutoff here
analogous to the quark mass in other applications.}
At larger coupling $N_{\rm CG}$ moderately rises
in the range that has been explored here.

The actual
cost to determine the running coupling at fixed error for $g^{-2}_{\rm SF}$
as discussed before hence \rf{seems to roughly grow
like $1/a^7$ in the continuum limit,
at least at the relatively weak coupling considered here.
This seems to be more optimistic than 
the quark mass dependence in some previous estimates,
for instance\footnote{See also the discussion in \cite{Karlrev}.}
in \cite{Gupta}.
One reason may be that our growth may be slightly underestimated
due to overhead on the small lattices.
Closer inspection reveals as another source of difference
that in our molecular dynamics steps
we are not forced to lower the step size $\Delta\tau$ at the rate usually
estimated for constant acceptance while lowering the quark mass.
Keeping $L$ however fixed in physical units means that $\beta$ rises
when $a/L$ becomes smaller which makes the gauge field smoother
at the same time. This could lead to smaller discretization errors
and partly be responsible for the
observed behaviour. The integration method \cite{SEXWEIN}
may in addition interfere with standard scaling on intermediate size
lattices.}

The vertical dotted line in Fig.~1 is located at $L/a=16$ and
points to $\mcost \approx 30$. This implies about 100 days
on 512 nodes for 3\% scale accuracy. Thus at least with the
next generation of APE1000 machines a serious continuum calculation
should be within reach.
For $L/a=12$, our most expensive lattices so far, we found
a slight preference for
PHMC. This conclusion holds also among
the larger couplings simulated, with the costs being approximately independent
of $g^2_{\rm SF}$ in the present range between $1.1$ and $1.8$.
All runs with $L/a=10$ have been performed employing HMC with SSOR\footnote{
Our even-odd preconditioned (P)HMC code is unsuitable
for this lattice size due to
machine topology.
} 
and reach up to
much larger couplings. Here the growth of $\mcost$ and $\dcost$ with 
$g^2_{\rm SF}$ becomes clearly visible but not dramatic, see Fig.~3.
\begin{figure}
\vspace{0.0cm}
\begin{center}
\psfig{file=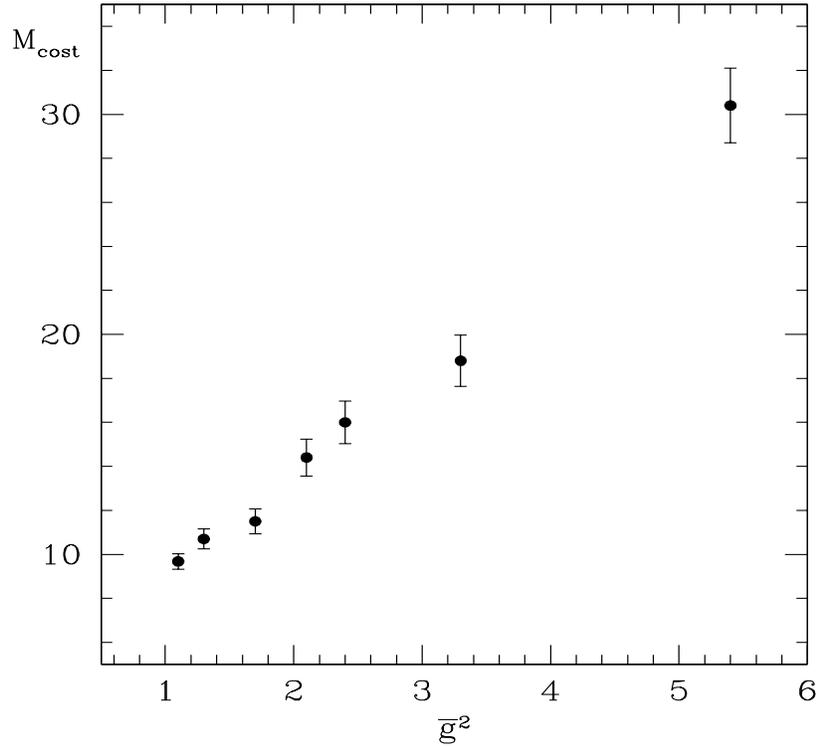, %
width=14cm,height=12cm}
\end{center}
\caption{Measured values of $\mcost$ for runs with HMC-SSOR 
on $L/a=10$ lattices
versus renormalized coupling.}
\end{figure}

With the MLM algorithm for improved fermions
we only have results for $L/a=5$
whose parameters are collected in Table 3. 
\begin{table}
\begin{center}
 \begin{tabular}{cccccccccccccc}
  \hline \\[-1.0ex]
  $\beta$ & $r_1$ & $n_1$ &  $n_2$ &  $n_3$ &  $n_4$ & $t_1$ &$t_2$&$t_3$
&$t_4$
  & $a_1$ &  $a_2$ & $a_3$ & $a_4$ \\[1.0ex]
  \hline \\[-1.0ex]
   8.39   &  4  & 15   & 31  & 63 & 255 & 1 & 6 & 6  & 10 & 0.313& 0.828& 0.966
   & 0.998 \\
   8.854  &  4  & 15   & 31  & 63 & 255 & 1 & 6 & 6  & 10 & 0.347& 0.835& 0.969
   & 0.999 \\
   9.40   &  3  & 11   & 23  & 63 & 255 & 1 & 8 & 10 & 10 & 0.355& 0.742& 0.923
   & 0.996 \\[1.0ex]
  \hline \\[-1.0ex]
 \end{tabular}
 \end{center}
\label{nmax}
\caption{
Parameters of the MLM update cycle, where $l=4$, $r_2=r_3=r_4=1$
and
$a_j$ denotes the acceptance rate at the level $j$.
}
\end{table}
As updates on level zero (gauge action only) we used heatbath sweeps and 
overrelaxation sweeps over certain sets of links. 
With probability $1/2$
we update either set A or set B.  
Set A consists of all spatial links of one randomly selected timeslice 
together with the 
temporal links at this timeslice in either positive or negative 
time direction.
Set B are all links with a randomly selected spatial direction plus 
the temporal boundary links.
In the case of set A we perform a heatbath sweep over all links followed
by five overrelaxation sweeps. 
In the case of set B we perform a heatbath sweep over the temporal boundary
links
followed by five overrelaxation sweeps over the spatial links of one direction.
In both cases  the order of the updating is exactly reversed with 
probability 1/2 to fulfill detailed balance. Given the large set of
free parameters, some of them had to be chosen rather ad hoc.
As we implemented and ran MLM
\rf{on PCs but not on APE100, to which our $\mcost$-values refer,}
we only quote the machine independent $\dcost$ here.
Together with results for the coupling, which were found consistent with 
HMC results, they are given in Table~4.
\begin{table}
 \vskip 0.2cm
\begin{center}
 \begin{tabular}{cccc}
\hline \\[-1.0ex]
   $\beta$ &   8.39   &  8.854  &   9.40 \\[1.0ex]
\hline \\[-1.0ex]
${g}^{2}_{\rm SF}$ & 1.1807(12)&1.0778(10)& 0.9767(16)\\[0.5ex]
$D_{\rm cost}$ & 290(6) & 287(6) & 372(15) \\[1.0ex]
\hline \\[-1.0ex]
\end{tabular}
\end{center}
\label{DcostMartin}
\caption{ 
$D_{\rm cost}$ for simulations with the multi-level algorithm of $5^4$ lattices
at various $\beta$-values.
 }
\end{table}

\section{Conclusions}
We have studied several simulation algorithms for the Schr\"odinger functional
of full QCD with two flavours of massless quarks.
Due to relatively high statistics on lattices up to $12^4$
we obtained precise information on integrated autocorrelation times.
Although our results are relatively close for the algorithms compared,
there is a slight advantage for the polynomial hybrid Monte Carlo
for our parameter range and observable.
For this numerical reason and for the expected advantages from its modified
sampling at larger coupling, we plan to focus on PHMC for our
coming runs with $L/a=16$.
In simulations with ordinary HMC we found even-odd
preconditioning of the {\it determinant} more efficient than
SSOR preconditioning the {\it solver} alone.

\noindent {\bf Acknowledgement} We would like to thank Rainer Sommer
for numerous discussions and for a critical reading of the manuscript.
We are grateful to DESY for allocating computer time on the APE machines at
DESY Zeuthen. 

\begin{appendix}
\section{Integrated autocorrelation time for functions of observables}

In this appendix we discuss a method to assess the 
effect of autocorrelations on the
statistical error of nonlinear
functions of simple expectation values.
We consider a number of observables in a statistical system, and
by $A_{\alpha}, \alpha=1,2,\ldots$ we denote their {\em exact}
mean values. For each
observable we have a chain of $N$ 
unbiased but (auto-)correlated Monte Carlo estimates $a_{\alpha}^i, i=1,\ldots, N$.
Assume that we want to estimate $F = f(A_{\alpha})$, where $f$ is an in principle
arbitrary function. A simple case arising in the context of reweighting is
the quotient $F=A_1/A_2$, while fit parameters extracted from a correlation function
at a sequence of separations would be a more complicated case.

The obvious estimator for $F$ is given by $f(\bar{a}_{\alpha})$,
where
\be
\bar{a}_{\alpha} = \frac1{N} \sum_{i=1}^N a_{\alpha}^i
\ee
are the ensemble-means for our simulation.
In a correct and equilibrated Monte Carlo we expect
\bea
\langle A_{\alpha} - \bar{a}_{\alpha} \rangle &=& 0, \\
\langle (A_{\alpha} - \bar{a}_{\alpha})^2 \rangle &=& \rO(1/N)
\eea
to hold, where the expectation values in this appendix mean the {\em average over}
an infinite number of identical {\em Monte Carlo simulations} of length $N$.
Loosely speaking, $A_{\alpha}$ and $\bar{a}_{\alpha}$ differ by $\rO(1/\sqrt{N})$
in an individual Monte Carlo run.

By Taylor expanding $f$ around the argument $A_{\alpha}$ 
we find
\bea
\langle F - f(\bar{a}_{\alpha}) \rangle &=& \rO(1/N), \\
\sigma^2 = \langle (F - f(\bar{a}_{\alpha}))^2 \rangle &=&  \rO(1/N).
\eea
The first line reveals the (in general unavoidable) bias of our estimator
which has to be suppressed\footnote{
In principle it is also possible to cancel the leading bias-term,
for instance by the jackknife method. 
} by large enough $N$. The statistical error
$\sigma$ of order $1/\sqrt{N}$ will be discussed in the following.
With the gradient vectors
\be
H_{\beta} = f_{| \beta}(A_{\alpha}), \quad
\bar{h}_{\beta} = f_{| \beta}(\bar{a}_{\alpha})
\ee
we define projected observables
\be
A_H = \sum_{\alpha} A_{\alpha} H_{\alpha}
\ee
and analogously with $\bar{h}_{\alpha}$ and for $\bar{a}_{\alpha}$.
Now we conclude that up to higher orders in $1/N$
we just need to know the variance of the projected observable,
\be
\sigma^2 \approx \langle (A_H -  \bar{a}_H)^2 \rangle \approx
\langle (A_{\bar{h}}  - \bar{a}_{\bar{h}} )^2 \rangle.
\ee
In practice, the projection 
can only be performed with $\bar{h}$, taken from the data, of course.

From here on one may proceed just like in the case of simple expectation
values. We may estimate the relevant autocorrelation function 
at separation $t$ for instance\footnote{
Less symmetrically, one might also subtract $\bar{a}_{\bar{h}}$ 
in each bracket.} as
\be
\Gamma(t) = \frac{1}{N-t} \sum_{i=1}^{N-t}
\left(a^i_{\bar{h}} - \frac{1}{N-t}\sum_{j=1}^{N-t} a^j_{\bar{h}}\right)
\left(a^{(i+t)}_{\bar{h}} - \frac{1}{N-t}\sum_{k=t+1}^{N} a^k_{\bar{h}}\right).
\ee
From it the error follows as
\be
\sigma^2 = \frac{\Gamma(0)}{N} 2 \tau_{int}
\ee
with
\be
\tau_{int} = \frac12 + \sum_{t=1}^W \frac{\Gamma(t)}{\Gamma(0)}.
\ee
The summation window $W$ is usually  chosen large enough that $\tau_{int}$
saturates to a constant within statistical errors. Often this can be achieved
by selfconsistently summing until $W/\tau_{int}$ reaches numbers like $5 \ldots 10$.
Below the role of $W$ will be discussed further.
In summary, the deviation of $2 \tau_{int}$ from one for the
projected observable describes the complete effect of 
(auto)correlations on the estimation of $F$.
Obviously, $\sigma$, $\Gamma$ and $\tau_{int}$ all depend on the function
$f(A_{\alpha})$ which has remained implicit in our notation.

We would like to conclude this appendix
by indicating the advantage of explicitly
summing $\Gamma$ compared to the jackknife binning procedure
which is often applied for the estimation of errors of
secondary quantities like best-fit parameters.
There one divides $N$ estimates into $N/B$ bins of length $B$.
The individual bins are treated as uncorrelated. The fact that
this is not exactly true leads to a systematic error
in the error estimate which is of order $\tau/B$ from the correlation
of neighbouring bins.
Here $\tau$
is a general scale of autocorrelation times involved. 
This is usually controlled by demanding a plateau of the
errors as the bin length is varied.
The statistical
uncertainty of the error estimate is of order $\sqrt{B/N}$.
These two errors have to be balanced at an optimal bin length,
which incidentally scales as $B \propto (N \tau^2)^{1/3}$.

If we sum $\Gamma$ up to separation $W$ (summation window), systematic
errors due to the neglected remainder are of order $\exp(-W/\tau)$.
The statistical error of the error estimate is expected to be
of order $\sqrt{W/N}$ from the number of independent windows.
In fact, Madras and Sokal quote  the formula \cite{MadrasSokal}
\be
\frac{\delta \tau_{\rm int}}{\tau_{\rm int}} = 
\sqrt{\frac{2(2 W +1)}{N}} \; ,
\ee
which follows if one approximates the required summed 4-point
autocorrelation function by its disconnected part which falls back
to the sum over $\Gamma$ entering into $\tau_{\rm int}$ itself.
In practice, these errors usually look very reasonable under repeated runs.
The conclusion is that the systematic errors for the ``summation method''
are much smaller, which, in balancing systematic with statistical
errors, leads to more accurate error estimates. Taking the idea of
balancing totally seriously, one would conclude that the ``error
of the error'' decays like $[1/N]^{1/3}$ with binning and with
$[\ln(N)/N]^{1/2}$ with the $\Gamma$--summation method.

\section{Tuning of the PHMC algorithm}

Here we summarize our strategy for tuning the free parameters
of the PHMC algorithm, in particular for the Schr\"odinger functional
at small volume or weak coupling.
We are interested in the error $\sigma_{N_{\rm corr}}$ of the 
estimate 
\be
\langle {\cal O} \rangle = \frac
{\langle {\cal O}\overline{W} \rangle_P}  
{\langle \overline{W} \rangle_P} ,
\label{eta_ave}
\ee
for some observable $\cal O$ (mainly $g_{\rm SF}^{-2}$ at present)
and $\overline{W}$ given in eq.~(\ref{DefWbar}).
All errors are
assessed by the method of the previous appendix.
Here the error can be decomposed as
\be
\sigma^2_{N_{\rm corr}} = 
\sigma^2_{P} + [\sigma^2_{\infty}-\sigma^2_{P}]
+ [\sigma^2_{N_{\rm corr}} - \sigma^2_{\infty}].
\label{DecompError}
\ee
Here $\sigma^2_{\infty}$ corresponds to $N_{\rm corr}=\infty$,
or equivalently to the use of the exact reweighting with $W$
as in eq.~(\ref{true_ave}), while $\sigma^2_{P}$ refers to
the simple mean $\langle {\cal O} \rangle_{P}$.
The second term in (\ref{DecompError}) reflects a contribution 
from ``ideal'' reweighting and hence from the imperfection of
the polynomial approximation, while the third one is due to our
non-ideal stochastic estimation of the correction.
Both would vanish for a perfect polynomial and
are naively proportional to $\delta^2$, the scale of polynomial errors.
The third term has a factor $1/N_{\rm corr}$ in addition.
We can roughly disentangle them by measuring 
$\langle {\cal O} \rangle_{P}$ and
$\langle {\cal O} \rangle$  and their errors
for at least two values of $N_{\rm corr}$.
The goal now is to take $N_{\rm corr} =1 \ldots 4$ and find
$\delta, \epsilon$ such that the reweighting part of the error
remains acceptable, less than 20\%, say. This is to be achieved
at the smallest possible value of the polynomial degree $n$.

In \cite{PHMC3} we found for 
$\beta$ between 5.4 and 6.8 on $8^3 \times 16$ the rule
$\delta \approx 0.01$ and $\epsilon \approx 2\lambda_{\rm min}$
to be very efficient, where $\lambda_{\rm min}$ is
the average smallest eigenvalue of $\hat{Q}^2$.
At the larger $\beta$-values of the present study we expect the
intrinsic fluctuations caused by the gauge field
to be smaller and correspondingly found a too large relative
contribution from the third term in (\ref{DecompError})
when the above tuning is employed. Instead we found it much more efficient
to attenuate this term with a smaller $\delta$. This turned out
to be possible essentially without enlarging $n$, i.e. we could allow
$\epsilon$ to grow even larger than $2 \lambda_{\rm min}$.
This is probably due to smaller fluctuations of the small
eigenvalues as well. In Table 5 and 6 we report simulation
parameters of the PHMC runs on $12^4$.
\rf{As to the choice of $\tilde{c}_0$ it is noted that one has
to ensure $\lambda_{\rm max} < 1$ for all configurations for
a numerically stable evaluation of (\ref{ordering}). On the other hand,
the efficiency is not very sensitive to the precise value of
$\langle \lambda_{\rm max} \rangle$ which we hence kept safely
below one.}
\begin{table}
\vspace{2mm}
\centering
\label{simul_identif}
\begin{tabular}{cccccc}
\hline \\[-1.0ex]
 set & $\beta$ & $N_{\rm corr}$ & $\epsilon$ & $n$ & $\delta$ \\[1.0ex]
\hline \\[-1.0ex]
  12d       &  $9.5$    & $4$ & $0.0050$   & $44$ & $0.0034$\\
  12e       &  $9.5$    & $2$ & $0.0050$   & $44$ & $0.0034$\\
  12g       &  $8.5$    & $4$ & $0.0044$   & $44$ & $0.0050$\\
  12h       &  $8.5$    & $2$ & $0.0069$   & $42$ & $0.0016$\\
  12j       &  $7.5$    & $4$ & $0.0050$   & $46$ & $0.0026$\\
  12k       &  $7.5$    & $2$ & $0.0050$   & $46$ & $0.0026$\\[1.0ex]
\hline \\[-1.0ex]
\end{tabular}
\caption{Parameters controlling the polynomial approximation 
to $\hat{Q}^{-2}$ for PHMC runs on $12^4$.}
\end{table}
\begin{table}
\vspace{2mm}
\centering
\label{simul_tuning}
\begin{tabular}{ccccc}
\hline \\[-1.0ex]
 set & $\langle \lambda_{\rm min} \rangle$ & $\langle \lambda_{\rm max} 
\rangle$ & $\tilde{c}_0$ &
$\sigma^2_{N_{\rm corr}}/\sigma^2_{P}$\\[1.0ex]
\hline \\[-1.0ex]
  12d &  $0.002477(15)$    & $0.8582(1)$ & 0.6738653 & $1.05$\\
  12e &  $0.002474(12)$    & $0.8582(1)$ & 0.6738653 & $1.10$\\
  12g &  $0.002284(14)$    & $0.8765(1)$ & 0.6686256 & $1.35$\\
  12h &  $0.002270(13)$    & $0.8765(1)$ & 0.6686256 & $1.14$\\
  12j &  $0.001869(14)$    & $0.8664(1)$ & 0.6477127 & $1.06$\\
  12k &  $0.001860(15)$    & $0.8667(1)$ & 0.6477127 & $1.03$\\[1.0ex]
\hline \\[-1.0ex]
\end{tabular}
\caption{Average smallest and largest
eigenvalue of $\hat{Q}^2$ (see eq.~(\ref{Qhat}) for $\tilde{c}_0$)
and ratio of errors in eq.~(\ref{DecompError}).
}
\end{table}

\end{appendix}


\newpage
\begin{thebibliography}{99}

\bibitem{REV1}
M.~L\"uscher,
Lectures given at Les Houches Summer School  1997,
hep-lat/9802029.


\bibitem{REV2}
R.~Sommer  [ALPHA Collaboration],
Lectures given at 36. Internationale Universit\"{a}tswochen
f\"{u}r Kern- und Teilchenphysik,
Schladming 1997,
hep-ph/9712218.


\bibitem{QUENCHEDDATA}
S.~Capitani, M.~L\"uscher, R.~Sommer and H.~Wittig  [ALPHA Collaboration],
Nucl.\ Phys.\  {\bf B544}, 669 (1999)
[hep-lat/9810063].



\bibitem{JURI}
J.~Rolf and U.~Wolff,
Nucl.\ Phys.\ Proc.\ Suppl.\  {\bf 83-84}, 899 (2000)
[hep-lat/9907007].

\bibitem{SFA}
M.~L\"uscher, R.~Sommer, P.~Weisz and U.~Wolff,
Nucl.\ Phys.\  {\bf B413}, 481 (1994)
[hep-lat/9309005].


\bibitem{NPIMPR}
K.~Jansen and R.~Sommer  [ALPHA Collaboration],
Nucl.\ Phys.\  {\bf B530}, 185 (1998)
[hep-lat/9803017].

\bibitem{PCAC}
M.~L\"uscher, S.~Sint, R.~Sommer and P.~Weisz,
Nucl.\ Phys.\  {\bf B478}, 365 (1996)
[hep-lat/9605038].



\bibitem{HMC1}
S.~Duane, A.~D.~Kennedy, B.~J.~Pendleton and D.~Roweth,
Phys.\ Lett.\  {\bf B195}, 216 (1987).


\bibitem{SEXWEIN}
J.~C.~Sexton and D.~H.~Weingarten,
Nucl.\ Phys.\  {\bf B380}, 665 (1992).

\bibitem{deFor}
P.~de Forcrand and T.~Takaishi,
Nucl.\ Phys.\ Proc.\ Suppl.\  {\bf 53}, 968 (1997)
[hep-lat/9608093].


\bibitem{PHMC1}
R.~Frezzotti and K.~Jansen,
Phys.\ Lett.\  {\bf B402}, 328 (1997)
[hep-lat/9702016].

\bibitem{MBML}
M.~L\"uscher,
Nucl.\ Phys.\  {\bf B418}, 637 (1994)
[hep-lat/9311007].

\bibitem{Beat}
B.~Jegerlehner,
Nucl.\ Phys.\  {\bf B465}, 487 (1996)
[hep-lat/9512001].

\bibitem{MLMH}
M.~Hasenbusch,
Phys.\ Rev.\  {\bf D59}, 054505 (1999)
[hep-lat/9807031].

\bibitem{Qdefinition}
M.~L\"uscher and P.~Weisz,
Nucl.\ Phys.\  {\bf B479}, 429 (1996)
[hep-lat/9606016].

\bibitem{Bernd}
B.~Gehrmann and U.~Wolff,
Nucl.\ Phys.\ Proc.\ Suppl.\  {\bf 83-84}, 801 (2000)
[hep-lat/9908003].

\bibitem{JaLi95}
K.~Jansen and C.~Liu,
Nucl.\ Phys.\  {\bf B453}, 375 (1995)
[hep-lat/9506020].

\bibitem{JaLi96}
K.~Jansen and C.~Liu,
Comput.\ Phys.\ Commun.\  {\bf 99}, 221 (1997)
[hep-lat/9603008].

\bibitem{ssor}
S.~Fischer, A.~Frommer, U.~Glassner, T.~Lippert, G.~Ritzenhofer and K.~Schilling,
Comput.\ Phys.\ Commun.\  {\bf 98}, 20 (1996)
[hep-lat/9602019].

\bibitem{ssor1}
N.~Eicker, W.~Bietenholz, A.~Frommer, H.~Hoeber, T.~Lippert and K.~Schilling,
Nucl.\ Phys.\ Proc.\ Suppl.\  {\bf 63}, 955 (1998)
[hep-lat/9709143].
%

%
\bibitem{ssor2}
M.~Guagnelli and J.~Heitger  [ALPHA Collaboration],
Comput.\ Phys.\ Commun.\  {\bf 130}, 12 (2000)
[hep-lat/9910024].

\bibitem{bicg}
A.~Frommer, V.~Hannemann, B.~Nockel, T.~Lippert and K.~Schilling,
Int.\ J.\ Mod.\ Phys.\  {\bf C5}, 1073 (1994)
[hep-lat/9404013].

\bibitem{PHMC2}
R.~Frezzotti and K.~Jansen,
Nucl.\ Phys.\  {\bf B555}, 395 (1999)
[hep-lat/9808011].

\bibitem{PHMC3}
R.~Frezzotti and K.~Jansen,
Nucl.\ Phys.\  {\bf B555}, 432 (1999)
[hep-lat/9808038].

\bibitem{BUNK}
B.~Bunk, S.~Elser, R.~Frezzotti and K.~Jansen,
Comput.\ Phys.\ Commun.\  {\bf 118}, 95 (1999)
[hep-lat/9805026].


\bibitem{NHMB}
A.~Borrelli, P.~de Forcrand and A.~Galli,
Nucl.\ Phys.\  {\bf B477}, 809 (1996)
[hep-lat/9602016].





\bibitem{Gupta}
R.~Gupta, A.~Patel, C.~F.~Baillie, G.~Guralnik, G.~W.~Kilcup and S.~R.~Sharpe,
Phys.\ Rev.\  {\bf D40}, 2072 (1989).

\bibitem{Karlrev}
K.~Jansen,
Nucl.\ Phys.\ Proc.\ Suppl.\  {\bf 53}, 127 (1997)
[hep-lat/9607051].

\bibitem{NF2}
ALPHA Collaboration, in preparation.

\bibitem{MadrasSokal}
N.~Madras and A.~D.~Sokal,
J.\ Stat.\ Phys.\ {\bf 21}, 109 (1988).


\end{thebibliography}
\end{document}